# Validation of Habitable Zone Limits with a New Radiative–Convective Equilibrium Climate Model

Jacob Haqq-Misra[1], Eric T. Wolf[1,2], and Ravi K. Kopparapu[3]
[1] Blue Marble Space, Seattle, WA, USA; jacob@bmsis.org
[2] Laboratory for Atmospheric and Space Physics, University of Colorado Boulder, Boulder, CO, USA
[3] NASA Goddard Space Flight Center, Greenbelt, MD, USA

## Abstract

The habitable zone (HZ) boundaries calculated with the `Clima` one-dimensional radiative–convective equilibrium (RCE) model have served as a standard reference for exoplanet target selection for three decades, but they have never been independently validated with a comparable model. We present `ExoColumn`, a new one-dimensional RCE model that couples an independently developed atmospheric column physics package to the `ExoRT` correlated-*k* radiative transfer core that is also used by the three-dimensional `ExoCAM` model. We calibrate `ExoColumn` to present-day conditions, which matches other RCE models to within a few kelvin through the troposphere. Line-by-line benchmarks show that the `ExoRT` and `Clima` radiation cores each reproduce the outgoing longwave flux to within a few percent, with the differences due to the spectral line databases used by each model. We follow the same inverse climate modeling method as previous studies have done to find inner edge moist- and runaway-greenhouse limits of 0.970 au and 0.966 au, and an outer-edge maximum-greenhouse limit of 1.611 au at a $CO_2$ partial pressure of 8.7 bar. We extend these calculations to 12 host stars spanning 2000–7200 K and to planets of 0.1, 1, and 5 Earth mass, and we provide updated HZ boundary coefficients for community use. The `ExoColumn` limits systematically fall closer to the host star than those of `Clima` but agree to within a few percent in orbital distance, which validates the robustness of the published `Clima` HZ boundaries.

## 1. Introduction

Earth is the only planet in the solar system with liquid water on its surface, and it is the only planet known to host life. The realization that Venus and Mars are hostile to life as we know it is relatively recent in the history of astronomy, which supplanted the belief that life itself was ubiquitous across all planets (see, e.g., D. Danielson & C. M. Graney 2025). The rocky planets Earth and Venus are approximately the same size, while Mars is slightly smaller, so why is Earth the only planet that sustains oceans and a biosphere?

The region around a star where life can survive has been termed the "habitable zone" (HZ). Estimates by S. H. Dole (1964) found an optimistic HZ between 0.725 and 1.24 au, due to the use of a simplified representation of a thin atmosphere with fixed surface albedo. Calculations by M. H. Hart (1979) were the first to consider the increasing luminosity of a host star during its main-sequence phase, which produced a HZ in which life could avoid a runaway greenhouse at the inner edge and runaway glaciation at the outer edge. M. H. Hart (1979) found a narrow HZ range of 0.95–1.01 au for the solar system, and a nonexistent HZ for stars later than K0, which would imply that habitable exoplanets may be rare and limited to Sun-like stars.

Calculations by J. F. Kasting et al. (1993) suggested that previous models lacked the negative stabilizing feedback on Earth's climate that arises from the carbonate-silicate cycle, in which the long-term abundance of atmospheric $CO_2$ is regulated by temperature-dependent weathering with crustal rocks (J. C. Walker et al. 1981). The J. F. Kasting et al. (1993) model gave a conservative estimate of the HZ as 0.95–1.67 au, with the inner edge at the moist greenhouse limit where water is lost to space and the outer edge at the maximum greenhouse limit for a $CO_2$ atmosphere. J. F. Kasting et al. (1993) also noted that the HZ is wider for K- and M-type stars due to their slower main-sequence evolution; however, later type stars are also prone to synchronous rotation. Updates to the J. F. Kasting et al. (1993) model by R. K. Kopparapu et al. (2013, 2014) revised the HZ for the solar system to 0.99–1.70 au, with Earth residing closer to the innermost edge. The key insight from these HZ studies is that tectonically active terrestrial planets with a means of recycling volatiles (e.g., by a carbonate-silicate cycle) can include self-regulating negative feedback that can maintain habitable conditions despite a brightening host star. If such processes operate on some terrestrial exoplanets, then this implies that habitable exoplanets may be common and may occur around various stellar types.

The value in HZ studies is their operationalization in the search for life. The boundaries of the J. F. Kasting et al. (1993) and R. K. Kopparapu et al. (2013, 2014) HZ provide constraints on the orbital distances at which a tectonically active planet could maintain surface liquid water around a given star, which can provide a basis for choosing specific targets to observe with large ground- and space-based telescopes. The HZ concept does not categorically rule out the possibility of life existing in other places (for example, on the icy moons of outer planets), and likewise it is also incorrect to presume that a planet within its host star's HZ must inevitably be habitable. Certainly, any presumption of "inhabitance" is a severe misapplication of the definition of the HZ.

The HZ concept has been widely utilized since its inception and has motivated numerous applications of the concept. A noncomprehensive list of examples includes HZ calculations for dry and land-dominated planets (Y. Abe et al. 2011); hydrogen-dominated greenhouse worlds beyond the classical HZ (e.g., R. Pierrehumbert & E. Gaidos 2011; R. M. Ramirez & L. Kaltenegger 2017); three-dimensional climate modeling of the inner-edge limits (e.g., J. Leconte et al. 2013; R. K. Kopparapu et al. 2016); occurrence rate estimates for habitable exoplanets (e.g., R. K. Kopparapu 2013; C. D. Dressing & D. Charbonneau 2015; S. Bryson et al. 2021); characterization of observed exoplanet systems (e.g., E. V. Quintana et al. 2014; G. Anglada-Escudé et al. 2016; M. Gillon et al. 2017; P. Robertson et al. 2026); and critiques of the HZ concept (A. Zsom 2015; W. B. Moore et al. 2017; E. Tasker et al. 2017).

The J. F. Kasting et al. (1993) and R. K. Kopparapu et al. (2013, 2014) lineage of models (hereafter referred to as `Clima`) continue to serve as reference HZ boundaries for guiding observations and for comparing against computationally expensive three-dimensional models. However, no independent validation study of these HZ limits have been conducted, using a comparable radiative–convective equilibrium (RCE) climate model. Although RCE models have a long history in the study of Earth's climate (e.g., S. Manabe & R. T. Wetherald 1967), the `Clima` lineage has been the only RCE model with the extended capabilities of simulating moist atmospheres near the inner edge of the HZ and dense-$CO_2$ atmospheres at the outer edge of the HZ. In this paper, we present the `ExoColumn` RCE model, which was designed to have the same functionality as `Clima`. `ExoColumn` contains an independently developed RCE column model, which is paired with the `ExoRT` correlated-*k* radiation package used by the `ExoCAM` three-dimensional model (E. T. Wolf et al. 2022).

The design of `ExoColumn` is intended as one component in a coupled hierarchy of modeling tools. Problems in exoplanet climate modeling span a wide range of parameters, which correspond to increased computational cost, and not all models are equally suited for all tasks. For example, one-dimensional models are ideal for rapid exploration of large parameter spaces, while three-dimensional models are needed for explicit simulation of atmospheric dynamics, clouds, and surface processes. Using a hierarchy of models works best when the models share components, as this allows any differences between the models to be attributed to the components that are not shared. Because `ExoColumn` depends upon `ExoRT` without modifying it, the radiative transfer applied in this one-dimensional model is the same code, using the same spectroscopic data, that `ExoCAM` applies in three dimensions. This coupling ensures that `ExoColumn` will continue to be developed alongside `ExoCAM`, as any improvements made to `ExoRT`, such as updated line lists, revised continuum treatments, additional absorbing species, or new spectral configurations, will propagate to `ExoColumn` without additional effort. Standalone one-dimensional models, by contrast, tend to receive spectroscopic updates or new greenhouse gas constituents on an ad hoc basis. `ExoColumn` is publicly available and is designed to be straightforward to compile and run, following the same open development model as `ExoCAM`.

Although the calculations presented here are for terrestrial planets, `ExoColumn` is a general-purpose column model that is not restricted to Earth-like atmospheres. The atmospheric composition is specified at runtime from the set of radiatively active gases carried by `ExoRT`, and the model computes the mean molecular weight and specific heat of the background atmosphere directly from that composition, so that $N_2$-, $CO_2$-, and $H_2$-dominated atmospheres are all treated consistently (e.g., R. Pierrehumbert & E. Gaidos 2011; R. M. Ramirez & L. Kaltenegger 2017). Other radiatively active gases available in `ExoRT` include $CH_4$, $C_2H_6$, CO, $NH_3$, $O_2$, and $O_3$. The same model can therefore be applied across classes of planets, from rocky planets within the HZ to hydrogen-rich sub-Neptunes and mini-Neptunes, which we intend to explore in future work.

In the present paper, we present a description of `ExoColumn`, which is intended as a validation of previous HZ studies with `Clima` and a demonstration of the model's capabilities for subsequent use. We first tune the model to present-day Earth conditions and compare the radiative transfer against line-by-line benchmarks (Section 2). We perform calculations for the inner (Section 3) and outer (Section 4) edges of the HZ, following the same model configuration as R. K. Kopparapu et al. (2013, 2014). We show HZ limits across spectral types (Section 5) and provide updated HZ boundary coefficients for community use.

## 2. The ExoColumn Model

`ExoColumn` is a one-dimensional RCE model written in Fortran as standalone software, which depends on `ExoRT` without any modifications. This gives `ExoColumn` the flexibility to incorporate any future `ExoRT` updates. `ExoColumn` calls the `ExoRT` radiation routine (`aerad_driver`) directly, using the radiative fluxes from `ExoRT` to calculate steady-state vertical profiles of temperature. The default configuration uses the n68equiv 68-band scheme and a grid of 70 vertical layers log-spaced from $p_{top}$ to the surface. The `ExoRT` correlated-*k* coefficients are generated with `HELIOS-K` (S. L. Grimm & K. Heng 2015) from the HITRAN-2016 line lists (I. E. Gordon et al. 2017) and are tabulated separately for each absorbing gas over 17 temperatures spanning 100–500 K in increments of 25 K and 61 pressures spanning $10^{-5}$–10 bar in increments of 0.1 in $\log_{10}p$, with eight Gauss points in each spectral band. The overlap between gases is evaluated at runtime by the equivalent-extinction method (D. S. Amundsen et al. 2017), so there is no restriction on the abundance of any individual gas. If a layer falls beyond the tabulated range, `ExoRT` holds the absorption coefficients fixed at the nearest tabulated temperature. The corresponding `Clima` tables span 100–600 K and $10^{-5}$–100 bar on a coarser eight-by-eight grid, with values beyond this range held fixed in the same way.

The default convection scheme is simplified Betts–Miller (A. K. Betts 1986; A. K. Betts & M. J. Miller 1986; D. M. W. Frierson 2007), which relaxes the temperature profile toward a moist adiabat and humidity toward a reference relative humidity ($RH_{init}$ = 0.7). Other convection schemes are also selectable: dry convective adjustment (S. Manabe & R. F. Strickler 1964), Manabe–Wetherald moist adjustment schemes (S. Manabe & R. T. Wetherald 1967), and the Zhang–McFarlane scheme (G. J. Zhang & N. A. McFarlane 1995). The water vapor saturation vapor pressure ($q_{sat}$) is calculated with the Clausius–Clapeyron relation. `ExoColumn` also implements a stratospheric "cold trap" intended to represent stratospheric drying due to the Brewer–Dobson cycle. The model defines the cold-point tropopause as the level of minimum saturation mixing ratio, so that an ascending parcel's water vapor is reduced to this value at the tropopause and conserved on its continued ascent into the stratosphere.

`ExoColumn` is initialized either from a cold start or from an input profile. In a cold start, the initial temperature is set by integrating a moist adiabat upward from the surface temperature $T_s$, capped at an isothermal stratosphere at $T_{\rm strato}$; water vapor is set to $\mathrm{RH_{init}}\, q_{\rm sat}$ in the troposphere (and dried in the stratosphere). All the dry gases are well-mixed, with an option to include a modern Earth ozone profile (A. A. Wing et al. 2018). The initial state can also be read from a stored profile. The temperature profile is advanced toward RCE by time-marching: at each step, `ExoRT` returns the longwave and shortwave radiative heating rates for the current profile. These are applied to all layer temperatures as well as the surface sensible and latent heat sources in the lowest layer. Convective adjustment then restores the column to a lapse-rate-stable and subsaturated state.

The surface is a 10 m slab ocean ($H_{\rm slab} = 4.1 \times 10^7\ \mathrm{J\,m^{-2}\,K^{-1}}$), with surface temperature calculated with semi-implicit Planck damping. First, the surface upward longwave emission $\sigma T_s^4$ is linearized about the current surface temperature ($T_s$) so that its Planck feedback $4\sigma T_s^3$ is treated implicitly as:

$$T_s \leftarrow T_s + \Delta t \frac{F_{\rm net}^{\rm srf} - \mathrm{LE} - \mathrm{SH}}{H_{\rm slab} + \Delta t\ 4\sigma T_s^3}, \qquad (1)$$

where $F_{\rm net}^{\rm srf}$ is the net surface radiative flux, LE is the latent heat flux, and SH is the sensible heat flux. This is intended to damp temperature overshoots and keep the slab energy budget stable at large time steps. The column is marched to equilibrium with an adaptive time step set by a Courant–Friedrichs–Lewy (CFL) stability limit. Each step is sized so that the largest radiative temperature tendency in the column produces a change of at most $\Delta T_{\rm target} = 1$ K, with the resulting step restricted to the range $\Delta t = [10^{-4}, 1]$ day. The model is considered converged when either of two criteria is met. *Radiative equilibrium convergence* occurs when the column maximum sum of longwave plus shortwave heating falls below 0.01 K $\mathrm{day^{-1}}$ and the net top-of-atmosphere (TOA) flux falls below 0.1 W $\mathrm{m^{-2}}$. *Profile stability convergence* occurs when the changes in time-mean profiles over successive 100 day windows fall below fixed thresholds ($\max|\Delta T| < 1$ K in the temperature profile, $|\Delta T_s| < 0.02$ K at the surface, and $\max|\Delta q| < 5 \times 10^{-4}$ kg $\mathrm{kg^{-1}}$ in water vapor) while the window-mean TOA flux stays below 0.1 W $\mathrm{m^{-2}}$.

`ExoColumn` has no separate boundary-layer scheme; instead, surface turbulent fluxes use a simplified Monin–Obukhov formulation (D. M. W. Frierson et al. 2006) that is paired with a surface-coupled mixed layer. The latent and sensible heat fluxes that appear in Equation (1) are expressed as

$$\mathrm{SH} = \rho_a\, c_p\, C\, U\, (\Theta_s - \Theta_a), \qquad (2)$$

$$\mathrm{LE} = \rho_a\, L(T_s)\, C\, U\, (q_s^* - q_a), \qquad (3)$$

where $\rho_a$ is air density at the lowest model level, $U$ is the near-surface wind speed (a fixed model parameter, default 5 m $\mathrm{s^{-1}}$), $C$ a dimensionless exchange coefficient, $\Theta_s = T_s$ and $\Theta_a = T_a\,(p_0/p_a)^{R_d/c_p}$ are the surface and near-surface potential temperatures, $q_s^* = q_{\rm sat}(T_s)$ is the saturation specific humidity at the surface, $q_a$ is the near-surface specific humidity, $c_p$ is the specific heat of dry air, $R_d$ is the gas constant of dry air, and $L(T_s)$ is the latent heat of vaporization evaluated at the surface temperature (which becomes the latent heat of sublimation below the triple point). The value of $C \approx 10^{-3}$ is calculated based on a fixed 10 m reference height in order to be resolution-independent and of Earth-like magnitude. These surface fluxes, and the slab energy budget of Equation (1), only operate when the column is time-marched to equilibrium, as in the present-day Earth calculation of Section 2.1. The HZ limits in Sections 3–5 are instead computed with the inverse method described next, in which the temperature and humidity profiles are prescribed and neither the surface turbulent fluxes nor the slab heat capacity enter the calculation.

`ExoColumn` includes several optional physics modules (disabled by default) that are needed to make a consistent comparison of the inner and outer edge HZ calculations with `Clima`. At the inner edge of the HZ, water vapor becomes a major atmospheric constituent and departs from ideal-gas behavior. We implement the nonideal moist pseudoadiabat described by J. F. Kasting (1988) and used in `Clima` as well as the IAPWS-95 nonideal equation of state for water (W. Wagner & A. Pruß 2002), which remains valid up to the critical point (647 K). We also included an option to replace the default MT_CKD water vapor continuum with the BPS continuum used in `Clima`. Toward the outer edge of the HZ, dense-$CO_2$ atmospheres develop due to reduced weathering rates. We implement an option for $CO_2$ condensation, in which the temperature profile is pinned to the $CO_2$ saturation curve (R. Span & W. Wagner 1996) wherever the dry adiabat would otherwise supersaturate in $CO_2$ (J. F. Kasting 1991). This option also includes a temperature-dependent $CO_2$ specific heat. We also include the capability to perform "inverse" climate modeling experiments in which a temperature profile is specified and the model is run a single step to calculate the corresponding solar flux needed to produce the given surface temperature, which allows us to perform the same set of HZ calculations as R. K. Kopparapu et al. (2013, 2014).

### 2.1. Earth Validation

We first validate `ExoColumn` against present-day Earth. The column is initialized from a cold start with a 1 bar background of $N_2$, $O_2$, and Ar, 400 ppm $CO_2$, 1.8 ppm $CH_4$, and the RCEMIP ozone profile (A. A. Wing et al. 2018). We calibrate the surface albedo so that the equilibrium surface temperature reaches the observed global-mean value of $T_s \approx 288$ K. This use of surface albedo as a "tuning parameter" follows the approaches of J. F. Kasting et al. (1993) and R. K. Kopparapu et al. (2013, 2014), with a physical justification that the tuned albedo accounts for the influence of partial cloud cover that cannot otherwise be represented in a one-dimensional column model. The calibrated surface albedo of 0.2736 allows `ExoColumn` to reach an equilibrium surface temperature $T_s = 287.84$ K, which converges in 2107 steps (≈1000 model days) to a top-of-atmosphere (TOA) flux imbalance below 0.1 W $\mathrm{m^{-2}}$. The resulting temperature profile, radiative fluxes, heating rates, and water vapor profiles are shown in Figure 1, alongside the `konrad` RCE model (L. Kluft et al. 2019) based on the `RRMTG` radiative transfer package (E. J. Mlawer et al. 1997), the `Clima` model, and the 1976 US Standard Atmosphere for comparison.

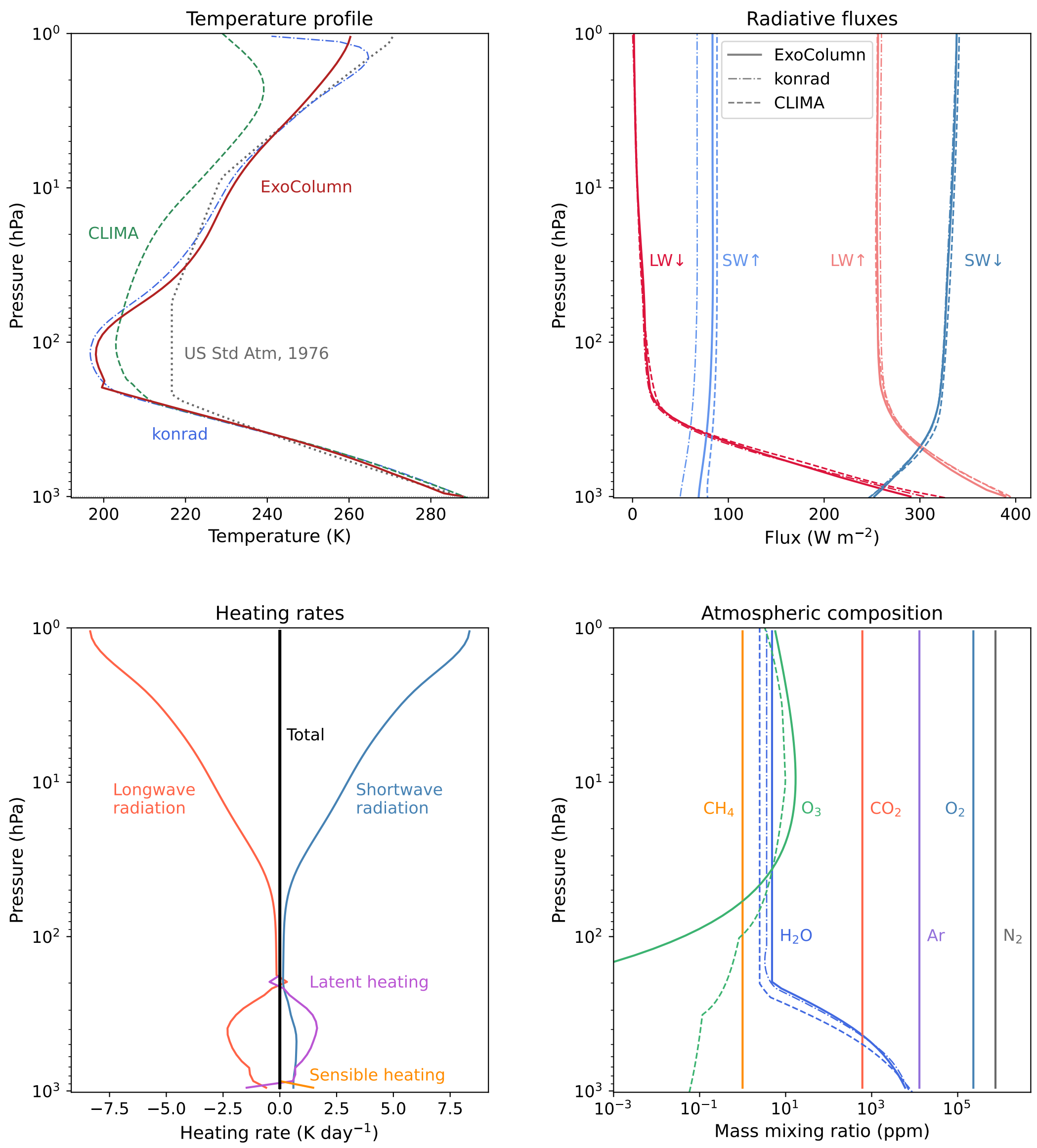

**Figure 1.** Present-day Earth validation computed with `ExoColumn`. The four panels show (top left) the vertical temperature profile, (top right) upward and downward longwave (LW) and shortwave (SW) radiative fluxes, (bottom left) heating rates, and (bottom right) atmospheric composition. Also shown for comparison are `konrad` (dashed–dotted), `Clima` (dashed), and the 1976 US Standard Atmosphere (dotted). The solid ozone curve is the RCEMIP analytic profile (A. A. Wing et al. 2018) used by `ExoColumn` and `konrad`. The dashed ozone curve is the US Standard Atmosphere profile used by `Clima`.

The `ExoColumn` temperature profile matches both `konrad` (calibrated surface albedo of 0.20) and `Clima` (calibrated surface albedo of 0.320) to within a few K through most of the troposphere. All models are consistent with the 1976 US Standard Atmosphere below the tropopause, with significant differences in the stratosphere; this is expected because the US Standard Atmosphere is based on observed averages. All three models share the same qualitative structure with a convecting troposphere at close to the moist adiabatic lapse rate, a temperature minimum near the tropopause, and a stratosphere in radiative equilibrium. The `ExoColumn` profile exhibits a sharp change in lapse rate (a temperature "kink") at the tropopause, where the convectively adjusted troposphere meets the radiative-equilibrium stratosphere. This abrupt lapse-rate change at the convective top is a property that appears in other RCE models with convective adjustment (see, e.g., I. M. Held 1982; J. Thuburn & G. C. Craig 2000) and not a numerical artifact. The `konrad` model produces a smoother tropopause because it applies a hard adjustment scheme to a prescribed moist lapse rate along with a cold-point tropopause coupling intended to produce a present-day Earth profile. The `Clima` model gives a tropopause at a lower altitude than either `ExoColumn` or `konrad`. The three models differ most in the stratosphere, where `ExoColumn` is the warmest and wettest, and the cold-point tropopause temperature minimum (198.1 K) is 1.4 K warmer than `konrad` (196.7 K) and 4.9 K colder than `Clima` (203.0 K).

Much of the difference in the stratosphere arises from the prescribed ozone profiles (Figure 1, bottom right), because shortwave absorption by ozone sets the stratospheric temperature inversion. `ExoColumn` and `konrad` use the identical

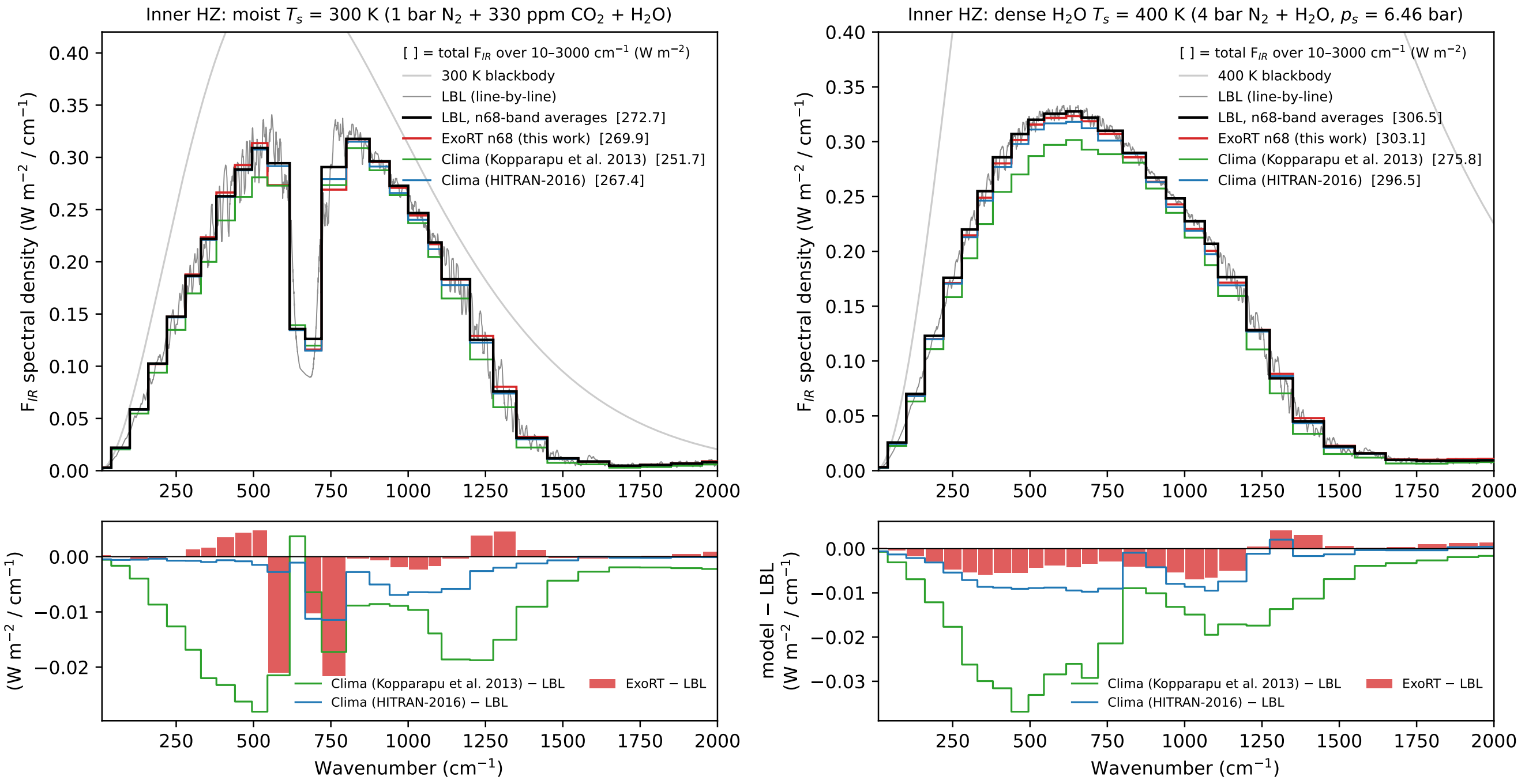

**Figure 2.** Comparison of the `ExoRT` (red) and `Clima` (green, blue) correlated-$k$ radiation cores against a line-by-line (LBL) reference (gray, black) for two moist atmospheric columns near the inner edge of the habitable zone. The left panel shows a warm $T_s = 300$ K column, and the right panel shows a dense-$H_2O$ column with $T_s = 400$ K that follows R. K. Kopparapu et al. (2013, Figure 2, right).

RCEMIP ozone profile, which is an analytic fit to a tropical-mean climatology (A. A. Wing et al. 2018) with a peak of 10.3 ppmv near 9.5 hPa and a column of 231 DU between the surface and the 1 hPa model top. `Clima` instead uses a US Standard Atmosphere ozone profile, represented as a piece-wise-linear function of $\log p$ that peaks at 6.0 ppmv at 10 hPa. The `Clima` profile carries more ozone in the upper troposphere and lower stratosphere (0.54 ppmv at 100 hPa, compared with 0.03 ppmv for RCEMIP) and a larger total column (273 DU to its 0.1 hPa model top), but it has less ozone at 3–30 hPa where the solar heating that drives the upper-stratospheric inversion peaks. To isolate this effect, we repeated the `ExoColumn` calculation with the surface albedo held at 0.2736 but with the `Clima` ozone profile in place of RCEMIP. The resulting `ExoColumn` stratosphere agrees with `Clima` to within 3.6 K (2 K on average) between 100 and 5 hPa, compared with differences of up to 13 K (9 K on average) with the RCEMIP profile, and the cold-point temperature (204.8 K) is within 2 K of `Clima`. The additional tropospheric ozone warms the surface to 291.4 K. A further calculation with no ozone gives a nearly isothermal stratosphere at 159–163 K above 30 hPa, which shows that the stratospheric temperature inversion in all three models is produced by ozone heating. The remaining differences in the stratosphere, particularly above 3 hPa where the model tops differ, and in the troposphere are attributable to differences between the underlying radiative transfer schemes.

### 2.2. Radiative-transfer Benchmarks

We next evaluate the `ExoRT` and `Clima` correlated-$k$ radiation cores for regimes relevant to the inner and outer edges of the HZ. We compare these both against independent line-by-line (LBL) radiative-transfer calculations. The LBL reference is built with a custom code that draws HITRAN line data and Voigt line profiles from the `RADIS` line-by-line code (E. Pannier & C. O. Laux 2019) and adds a sub-Lorentzian $\chi$-factor (M. Y. Perrin & J. M. Hartmann 1989) for the $CO_2$ far line wings as well as HITRAN-2024 $CO_2$–$CO_2$ collision-induced absorption. Figure 2 compares the outgoing longwave radiative flux ($F_{\rm IR}$) of `ExoRT`, `Clima`, and the LBL reference for moist atmospheres near the inner edge of the HZ, which includes a moist column at $T_s = 300$ K (left, with 1 bar $N_2$ and 330 ppm $CO_2$) and the dense-$H_2O$ column at $T_s = 400$ K (right, with a 4 bar $N_2$ background and $p_s = 6.46$ bar) used by R. K. Kopparapu et al. (2013, Figure 2). Figure 3 shows the same comparison for dense-$CO_2$ atmospheres near the outer edge, which includes a maximum-greenhouse column at $T_s = 273$ K (left, with 8.87 bar $CO_2$ and 1 bar $N_2$) and the dry early Mars column at $T_s = 250$ K (right, Mars-mass planet, 2 bar of 95% $CO_2$/5% $N_2$) used by R. K. Kopparapu et al. (2013, Figure 1). The early Mars column (Figure 3, right) also shows the corresponding `SMART` line-by-line spectrum as well as a `SOCRATES` calculation for the same column (J. M. Edwards & A. Slingo 1996; S. D. Guzewich et al. 2021).

Note that the `Clima` results shown in Figure 2 (both panels) and Figure 3 (left panel) are based on calculations we performed with the public version of `Clima`, whereas Figure 3 (right panel) shows the actual data used in the published figure of R. K. Kopparapu et al. (2013). (This is the reason that we do not show a corresponding HITRAN-2016 case as in the other panels.) The published version of `Clima` gives a total integrated flux of about 4 W m$^{-2}$ less than the value published by R. K. Kopparapu et al. (2013). We were unable to obtain an exact copy of the code that was used in the study by R. K. Kopparapu et al. (2013), but these differences between the `Clima` model generations are small.

Both `ExoRT` and `Clima` agree with the LBL reference to within a few percent. The differences are almost entirely due to the different generations of spectroscopic line data used in each model. The LBL reference itself is based on the HITRAN-2020 line list (I. E. Gordon et al. 2022), although differences from HITRAN-2016 in the bands that dominate the outgoing longwave radiative flux are less than 1%. The `ExoRT` n68equiv *k*-coefficients are computed from the HITRAN-2016 line lists (I. E. Gordon et al. 2017) for $H_2O$, $CO_2$, and $CH_4$. The `Clima` *k*-coefficients of R. K. Kopparapu et al. (2013) instead use the HITRAN-2008 line lists (L. S. Rothman et al. 2009), supplemented by the high-temperature HITEMP-2010 database (L. S. Rothman et al. 2010) for near-infrared $H_2O$ bands that dominate the absorbed stellar flux near the inner edge. Note that a more recent version of `Clima` (M. R. Vidaurri et al. 2022) has implemented new *k*-coefficients from HITRAN-2016 that remove most of the differences with `ExoRT` (blue curves in Figures 2 and 3). We show the 2013-era `Clima` results in this paper to remain consistent with the HZ results reported by R. K. Kopparapu et al. (2013, 2014). Differences between `ExoColumn` and `Clima` occur in the pure-rotational $H_2O$ band in the far-infrared ($\lesssim$600 cm$^{-1}$, $\gtrsim$17 $\mu$m; +8.6 W m$^{-2}$) and across the 750–1250 cm$^{-1}$ window (8–13 $\mu$m; +5.6 W m$^{-2}$). These properties of the underlying radiative transfer core are the primary source of differences between the `Clima` HZ calculations by R. K. Kopparapu et al. (2013, 2014) and those described next with `ExoColumn`.

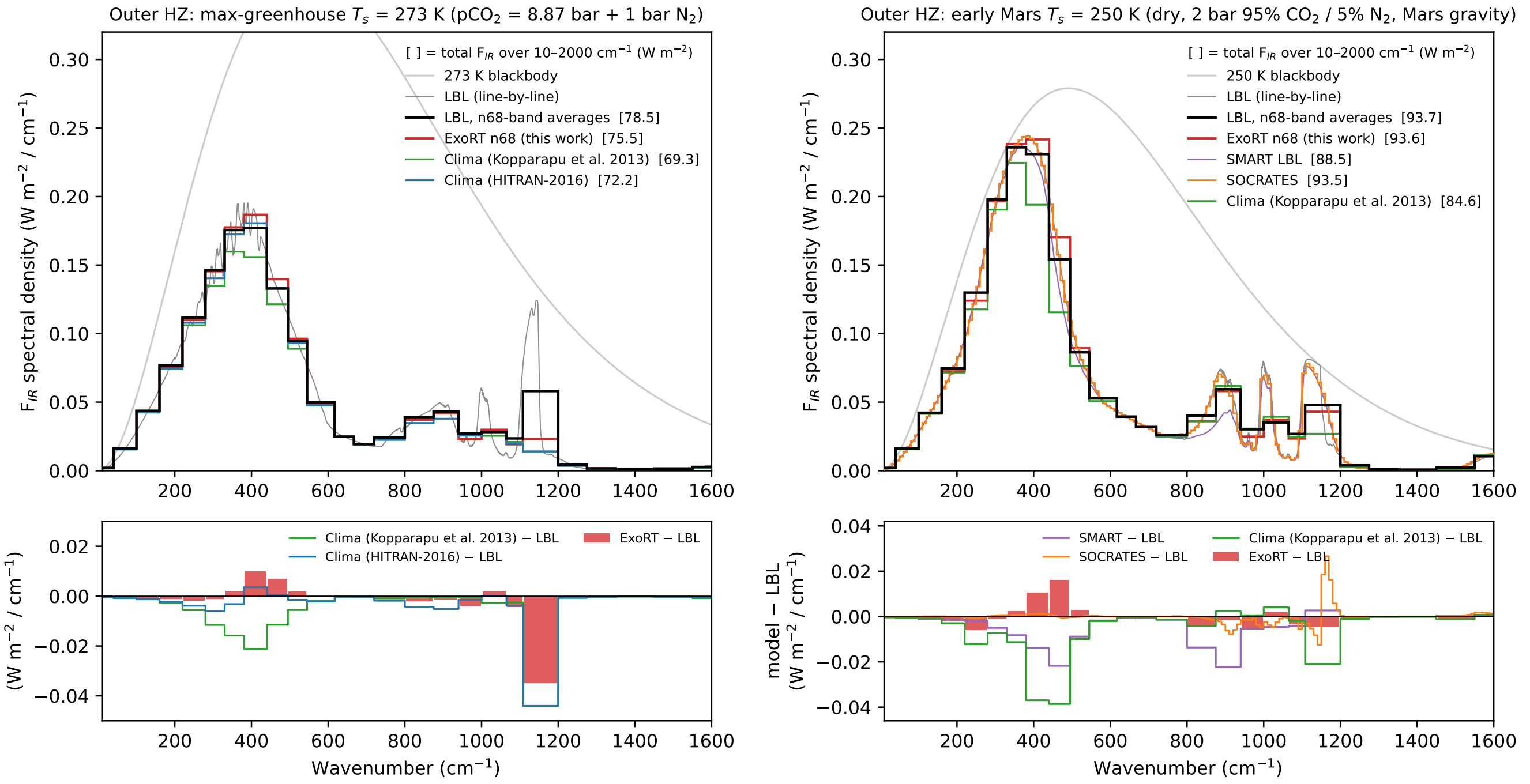

**Figure 3.** Comparison of the `ExoRT` (red) and `Clima` (green, blue) correlated-*k* radiation cores against a line-by-line (LBL) reference (gray, black) for two dense-$CO_2$ columns near the outer edge of the habitable zone. The left panel shows a $CO_2$-dominated column near the outer edge of the HZ. The right panel shows an early Mars column that follows R. K. Kopparapu et al. (2013, Figure 1), which also includes the corresponding `SMART` line-by-line spectrum (purple) and a `SOCRATES` calculation (orange).

The additional independent calculations for the dry early Mars column (Figure 3, right) provide further insight into the reasons for model discrepancies. This shows `ExoRT` with a net flux of 93.65 W m$^{-2}$ in close agreement with our own LBL reference (93.73 W m$^{-2}$) and the `SOCRATES` radiative transfer code (93.50 W m$^{-2}$). Note that `SOCRATES` and the LBL reference closely match below 750 cm$^{-1}$. At higher wavenumbers, the HITRAN-2024 $CO_2$–$CO_2$ collision-induced absorption tabulation carries no data. This is where the majority of the difference exists between `SMART`, `SOCRATES`, and the LBL reference. The model differences between 750 and 1120 cm$^{-1}$ thus appear to be due to the differences in the underlying $CO_2$–$CO_2$ collision-induced absorption.

## 3. The Inner Edge of the Habitable Zone

We calculate the inner edge of the HZ for Earth following the inverse climate modeling method of J. F. Kasting (1988) and R. K. Kopparapu et al. (2013). Instead of time-marching each case to equilibrium, we prescribe a surface temperature $T_s$ and build a vertical profile for a fully water-saturated column. We integrate upward from $T_s$ with a nonideal moist pseudoadiabat (J. F. Kasting 1988) and the IAPWS-95 steam equation of state (W. Wagner & A. Pruß 2002), with an isothermal stratosphere of $T_{strato}$ = 200 K. The `ExoRT` radiation core is then called a single step to return the net outgoing infrared radiative flux, $F_{IR}$, and the net incident solar radiative flux, $F_{SOL}$, for the prescribed column. (Note that $F_{SOL}$ is computed with a six-point Gauss–Legendre hemispheric average over the solar zenith angle, following R. K. Kopparapu et al. 2013.) The effective stellar flux that would sustain each column can then be calculated as $S_{eff} = F_{IR}/F_{SOL}$, so that $S_{eff} > 1$ denotes a planet receiving greater insolation than present-day Earth. These calculations assume a background of 1 bar of $N_2$ with 330 ppm $CO_2$, with no $O_2$, Ar, $O_3$, or $CH_4$, following the inner-edge configuration of R. K. Kopparapu et al. (2013). The $CO_2$ mixing ratio is held fixed relative to the total (moist) atmosphere, which is also the `Clima` convention. The surface pressure is the sum of the $N_2$ partial pressure and the saturation vapor pressure of water at $T_s$, and the water vapor profile is prescribed as saturated rather than computed from surface evaporation, so the latent and sensible heat fluxes of Equations (2)–(3) do not enter the inverse calculation. We assume a fixed surface albedo of 0.2736, the value calibrated in Section 2.1, so that

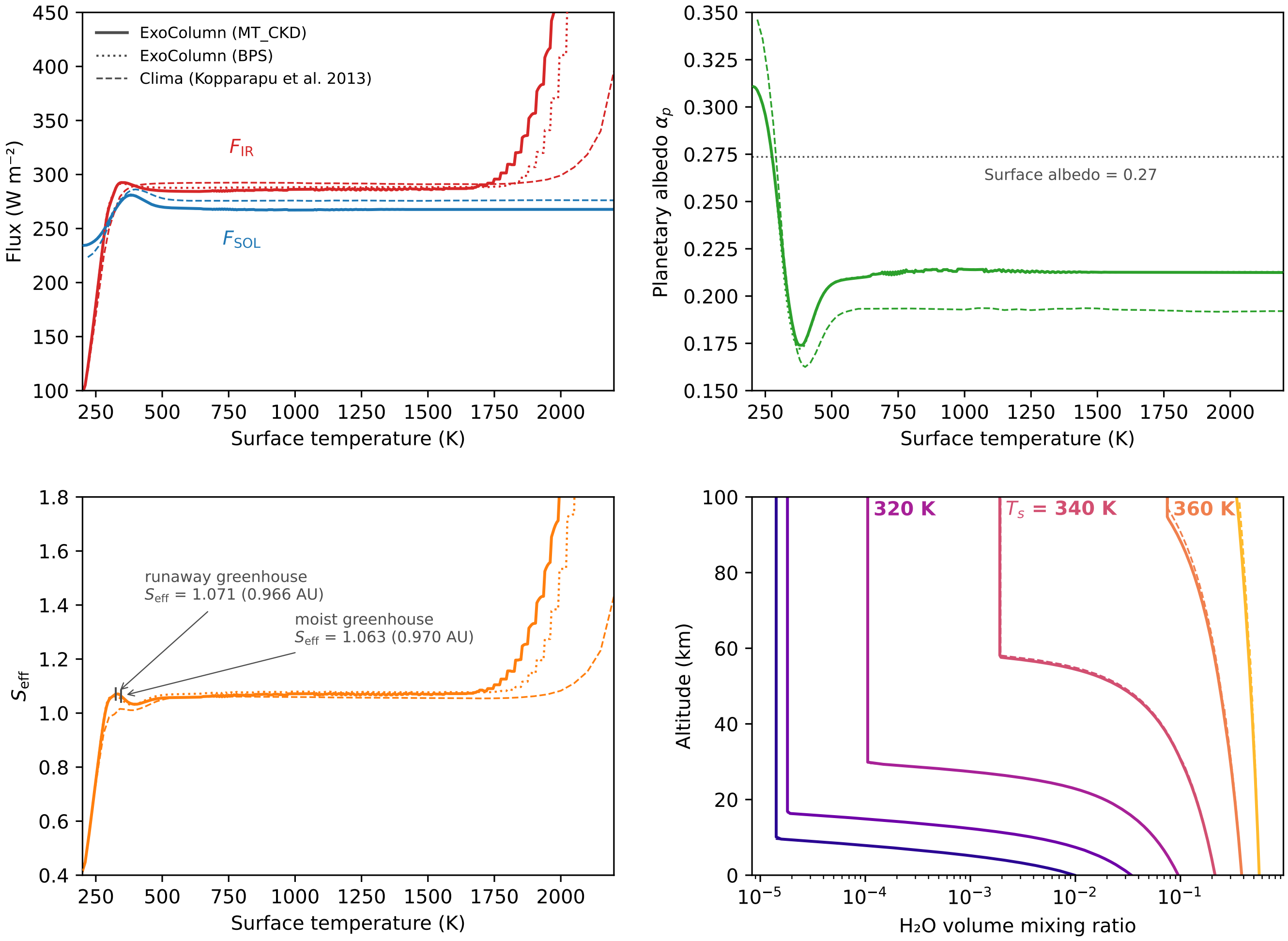

**Figure 4.** Inner edge of the HZ for a Sun-like star. Calculations with `ExoColumn` are shown using the default MT_CKD continuum (solid) and the BPS continuum (dotted), alongside `Clima` calculations with the BPS continuum for comparison (dashed). Panels show (top left) net outgoing infrared flux and net incident solar flux, (top right) planetary albedo, (bottom left) effective solar flux, and (bottom right) sample nonideal pseudoadiabatic water vapor profiles. The moist- and runaway-greenhouse limits marked in the bottom-left panel are $S_{eff} = 1.063$ (0.970 au) and 1.071 (0.966 au) for the MT_CKD continuum. The corresponding BPS-continuum limits are 1.052 (0.975 au) and 1.062 (0.970 au), and the `Clima` limits are 1.016 (0.992 au) and 1.060 (0.971 au).

`ExoColumn` reproduces the present-day Earth surface temperature. This follows the same approach as J. F. Kasting et al. (1993) and R. K. Kopparapu et al. (2013) by using the Earth-tuned albedo value for all HZ calculations. We perform the set of calculations by increasing $T_s$ upward from present-day values to the point where the column transitions into a moist and then runaway greenhouse state.

The results of these `ExoColumn` calculations for the inner edge of the HZ are shown in Figure 4, following the same format as R. K. Kopparapu et al. (2013, Figure 3). Values from `Clima` are also shown as well as a version of `ExoColumn` using the BPS water vapor continuum (used by `Clima`) instead of the default MT_CKD continuum. Figure 4 (top left) shows that $F_{IR}$ and $F_{SOL}$ steadily increase as the surface warms and water vapor comes to dominate the atmosphere, with $F_{IR}$ peaking near the Simpson–Nakajima radiation limit (S. Nakajima et al. 1992) of ≈295 W m$^{-2}$ (1.3% above the R. K. Kopparapu et al. 2013 value of 291 W m$^{-2}$) as most of the atmosphere becomes opaque to infrared radiation. (Note that C. Goldblatt et al. 2013 obtained a lower thermal limit of 282 W m$^{-2}$ using a LBL spectral resolution. Note also that R. T. Pierrehumbert 1995 observed that while local regions can reach runaway limit, dry subtropical regions act as "radiator fins" to stabilize the global climate.) $F_{SOL}$ likewise increases due to greater near-infrared absorption but then reaches a plateau as Rayleigh scattering intensifies. Further increases in $F_{IR}$ occur near 2000 K as the surface begins to radiate at visible and near-infrared wavelengths.

This behavior is shown in the changes of planetary albedo (calculated as the ratio of the upward and downward solar fluxes) with surface temperature (Figure 4, top right). The minimum value of planetary albedo corresponds to the maximum value in $F_{IR}$, which both then flatten out at higher surface temperatures. Note that the `ExoColumn` curves in Figure 4 retain a small step-like structure, which is a vertical-resolution sampling artifact as the tropopause kink crosses model levels. This effect is minimized by using 200 vertical levels in these `ExoColumn` inner HZ calculations.

The first inner HZ limit is the moist greenhouse limit, which is the point at which the stratospheric $H_2O$ mixing ratio reaches $3 \times 10^{-3}$, above which water is efficiently photodissociated and hydrogen escapes to space (J. F. Kasting 1988; R. K. Kopparapu et al. 2013). This change in water vapor content occurs between $T_s = 320$ K and 340 K, with over an order of magnitude increase in stratospheric water vapor (Figure 4, bottom right). This moist greenhouse limit occurs in the `ExoColumn` results at $S_{eff} = 1.063$ (Figure 4, bottom left), which corresponds to an orbital distance of $d = 1/\sqrt{S_{\mathrm{eff}}} = 0.970$ au. For comparison, the ExoColumn BPS-continuum moist greenhouse limit is at $S_{\mathrm{eff}} = 1.052$ (0.975 au), and the Clima limit is $S_{\mathrm{eff}} = 1.016$ (0.992 au).

The second inner HZ limit is the runaway greenhouse limit, which is the maximum stellar flux for which a planet with a surface ocean can remain in equilibrium. A planet receiving a greater flux would warm without limit until its oceans were completely evaporated. The runaway greenhouse limit occurs in the ExoColumn results at $S_{\mathrm{eff}} = 1.071$ (Figure 4, bottom left), which corresponds to an orbital distance of $d = 1/\sqrt{S_{\mathrm{eff}}} = 0.966$ au. This is the value of $S_{\mathrm{eff}}$ where the model first reaches the Simpson–Nakajima radiation limit (S. Nakajima et al. 1992), which occurs at $T_s \approx 325$ K. At this point the saturation vapor pressure is only ≈0.13 bar, or about 1.4 m of liquid water spread over the surface of an Earth-mass planet, so the runaway greenhouse limit is reached long before the ocean is depleted. Complete evaporation of the ocean occurs in the model only at the critical point of water (647 K), above which no liquid surface can exist and the full water inventory (270 bar, equivalent to Earth's oceans) is placed in the atmosphere. The runaway greenhouse limit is therefore independent of the size of the water inventory, provided that the planet holds more water than the ≈0.13 bar that is in the atmosphere at the Simpson–Nakajima limit. For comparison, the ExoColumn BPS-continuum runaway greenhouse limit is at $S_{\mathrm{eff}} = 1.062$ (0.970 au), and the Clima limit is $S_{\mathrm{eff}} = 1.060$ (0.971 au).

At surface temperatures near 2000 K, the thermal emission of the surface and the deep atmosphere peaks near 1.5 $\mu$m and escapes to space through the windows between the near-infrared $H_2O$ bands, so $F_{\mathrm{IR}}$ depends on the weak line and continuum opacity within those windows. The rise in $F_{\mathrm{IR}}$ begins earlier in ExoColumn than in Clima: $F_{\mathrm{IR}}$ first exceeds 300 W m$^{-2}$ at $T_s \approx 1775$ K in ExoColumn but at $T_s \approx 2050$ K in Clima. Both models prescribe the temperature profile with the same nonideal pseudoadiabat and apply no convective adjustment, so the divergence arises from the treatment of spectroscopic data at high temperature. About 100 K of the offset is due to the $H_2O$ continuum, as ExoColumn with the BPS continuum rises at $T_s \approx 1880$ K (dotted curves in Figure 4). The remainder is due to the near-infrared $H_2O$ line data, as the Clima $k$-coefficients include HITEMP-2010 hot lines (L. S. Rothman et al. 2010) that partially fill these windows at high temperature (Section 2.2). The limited temperature range of the $k$-coefficient tables does not explain the difference. Layers in the lower atmosphere reach well above the table limits in this regime (500 K for ExoRT and 600 K for Clima; Section 2), but when we lowered the ExoRT limit to 450 K and 400 K, the rise was delayed by only 25 and 55 K, and $F_{\mathrm{IR}}$ at 2000 K fell from 493 to 435 and 374 W m$^{-2}$. A higher table limit would therefore, if anything, make the ExoColumn rise begin earlier. Because both models hold their opacities fixed above the table limits, neither is quantitatively reliable above $T_s \approx 1750$ K, and the postrunaway branch would require $k$-coefficients computed from high-temperature line lists. This does not affect the HZ limits, which are set at $T_s < 350$ K, where every layer lies within the tabulated range of both models.

The offsets between the ExoColumn and Clima model are modest and almost entirely due to the different correlated-$k$ radiation cores used by each model (see Section 2.2). Our experiments included an implementation of the BPS water vapor continuum to more directly compare with Clima, but this only accounts for a fraction of the differences between the model results. Nevertheless, these ExoColumn results give moist and runaway greenhouse limits with HZ distances that agree with those published by R. K. Kopparapu et al. (2013) within ∼2%.

## 4. The Outer Edge of the Habitable Zone

We calculate the outer edge of the HZ as the maximum greenhouse limit for a $CO_2$-dominated atmosphere, following the inverse modeling method of J. F. Kasting et al. (1993) and R. K. Kopparapu et al. (2013). Here the surface temperature is held fixed at 273 K over a 1 bar $N_2$ background, and the $CO_2$ partial pressure is swept from 1 bar up to 34.7 bar. For each $CO_2$ partial pressure, we build a water-saturated moist adiabat in which the temperature profile is pinned to the $CO_2$ saturation curve wherever the dry adiabat would otherwise supersaturate (i.e., $CO_2$ condensation; J. F. Kasting 1991, with the R. Span & W. Wagner 1996 saturation curve). The $CO_2$ contribution to the specific heat is evaluated at the local temperature, and the stratosphere is isothermal at 154 K. As in the inner-edge calculation, the ExoRT radiative core is then called a single step to return $F_{\mathrm{IR}}$ and $F_{\mathrm{SOL}}$, with the ratio giving $S_{\mathrm{eff}}$. The maximum-greenhouse limit, denoting the outer edge of the HZ, is the minimum value of $S_{\mathrm{eff}}$ over the range of $CO_2$ partial pressures.

The results of the ExoColumn calculations for the outer edge of the HZ are shown in Figure 5, following the same three-panel format as R. K. Kopparapu et al. (2013, Figure 5). Values from Clima are also shown for comparison. Figure 5 (top) shows that $F_{\mathrm{IR}}$ and $F_{\mathrm{SOL}}$ decrease as the $CO_2$ partial pressure increases, which indicates the increasing greenhouse effect of $CO_2$. Beyond ∼10 bar, $F_{\mathrm{IR}}$ reaches a floor of ≈72 W m$^{-2}$ as the thick $CO_2$ atmosphere becomes opaque at infrared wavelengths, while $F_{\mathrm{SOL}}$ continues to decrease as Rayleigh scattering increases. This corresponds to a high planetary albedo as $CO_2$ partial pressure increases (Figure 5, middle). The competing tendencies of $F_{\mathrm{IR}}$ and $F_{\mathrm{SOL}}$ lead to a minimum $S_{\mathrm{eff}} = 0.385$ (1.611 au) at a $CO_2$ partial pressure of 8.68 bar. This is the maximum greenhouse limit, beyond which $S_{\mathrm{eff}}$ decreases as the cooling by Rayleigh scattering dominates any increased greenhouse warming from additional $CO_2$.

The Clima maximum greenhouse limit calculated by R. K. Kopparapu et al. (2013) is $S_{\mathrm{eff}} = 0.343$ (1.70 au) at a $CO_2$ partial pressure of ∼8 bar. The ExoColumn limit is thus more conservative, which places the outer edge closer to the star by ∼6%. This difference is driven by the dense-$CO_2$ offsets in the two correlated-$k$ radiation cores discussed in Section 2.2. We also note that the ExoColumn calculations above 10 bar are performed by clamping the correlated-$k$ lookup at the top of the ExoRT pressure table. This freezes the pressure broadening at its 10 bar value, while the collision-induced absorption and continuum opacities continue to use the actual pressure.

## 5. Habitable Zone Limits across Stellar Spectral Types

We now proceed to calculate the boundaries of the HZ for main-sequence stars, following the method first performed by J. F. Kasting et al. (1993) and then by R. K. Kopparapu et al. (2013) using versions of Clima. We compute the HZ limits with `ExoColumn` for 12 host stars spanning spectral types F (7200 K) through G (5780 K, the Sun), K (4000–4800 K), and M (2000–3700 K), using BT-Settl spectral energy distributions (SEDs; F. Allard et al. 2012).

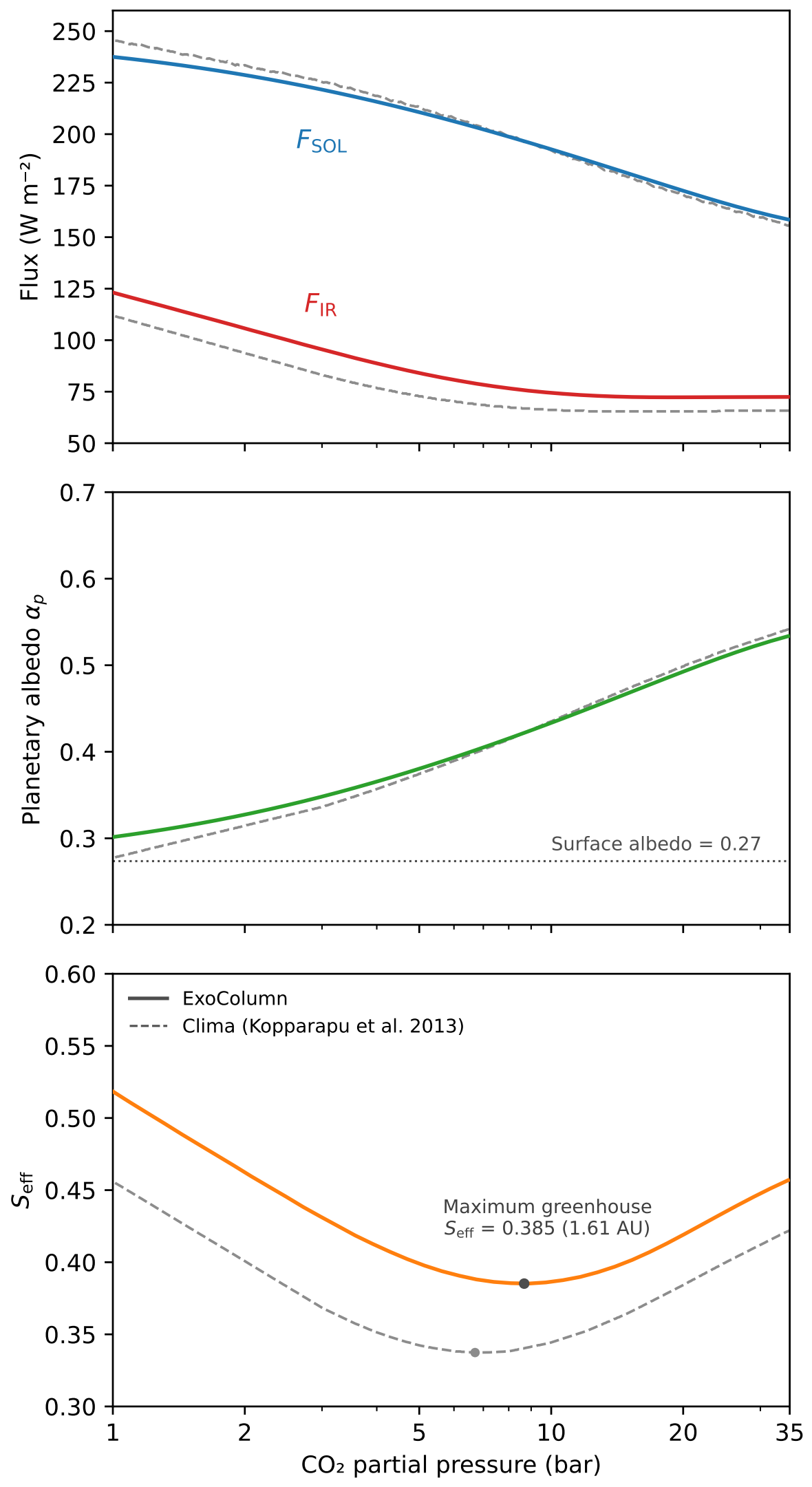

**Figure 5.** Outer edge of the HZ for a Sun-like star. Calculations with `ExoColumn` (solid) are shown alongside values from `Clima` (dashed). Panels show (top) net outgoing infrared flux and net incident solar flux, (middle) planetary albedo, and (bottom) effective solar flux. The maximum greenhouse limit marked on the bottom panel is $S_{eff} = 0.385$ (1.611 au) at a $CO_2$ partial pressure of 8.68 bar. The corresponding `Clima` value is $S_{eff} = 0.343$ (1.70 au) at ~8 bar.

Calculations of the inner and outer HZ edge for all 12 host stars are shown in Figure 6, following the four-panel format as R. K. Kopparapu et al. (2013, Figure 6). Planetary albedo (top left for inner edge, bottom left for outer edge) is higher for F-dwarf host stars and lowest for M-dwarf host stars, which occurs for two reasons. First is that the stellar spectral energy distribution of F-type stars is shifted toward shorter wavelengths, which increases the Rayleigh scattering of incident light. Second is that water vapor and $CO_2$ absorb more strongly at near-infrared than visible wavelengths, which results in M-dwarf planets absorbing a larger fraction of incident light. These effects are most pronounced for dense atmospheres, such as water-vapor-rich atmospheres near the inner edge or dense-$CO_2$ atmospheres near the outer edge.

These HZ boundaries are summarized in Figure 7, following the two-panel format of R. K. Kopparapu et al. (2013, Figure 7). Three boundaries are drawn in each panel: the runaway greenhouse and moist greenhouse limits (defining the inner edge) and the maximum greenhouse limit (defining the outer edge). The region between these curves is the conservative HZ. The two inner edges are nearly coincident over most of the range and deviate significantly only for the F-type host at 7200 K. The top panel shows $S_{eff}$ the HZ boundaries as a function of stellar effective temperature, and the bottom panel converts these limits to orbital distance as a function of stellar mass, using the I. Baraffe et al. (1998) 5 Gyr solar-metallicity relationships for stellar mass, luminosity, and effective temperature. The dashed curves in Figure 7 show the corresponding `Clima` boundaries (R. K. Kopparapu et al. 2013). The `ExoColumn` inner and outer limits both shift inward toward the star across all stellar types. The primary driver of this shift is the different correlated-$k$ radiation cores used in the two models (see Section 2.2). The net width of the HZ remains approximately the same as `Clima`, and the differences between the two models are only of a few percent when shown as a function of orbital distance (Figure 7, bottom panel). These results provide encouraging validation that the previously published `Clima` calculations of the HZ are robust when assessed with an independent model.

We fit the resulting HZ boundaries, following the quartic form of R. K. Kopparapu et al. (2013),

$$S_{eff}(T_{eff}) = S_{eff,\odot} + a\,T_\star + b\,T_\star^2 + c\,T_\star^3 + d\,T_\star^4, \qquad (4)$$

with $T_\star = T_{eff} - 5780$ K. This allows the associated HZ distances to be calculated as

$$d = \left(\frac{L/L_\odot}{S_{eff}}\right)^{0.5} \text{au}, \qquad (5)$$

where $L/L_\odot$ is the relative luminosity of the star to the Sun. This expression fits the `ExoColumn` model results with a rms residual of $\sim 10^{-3} S_\odot$ over 2000–7200 K. The resulting coefficients are listed in Table 1, with $S_{eff,\odot} = 1.07$, 1.06, and 0.39 for the runaway-, moist-, and maximum-greenhouse limits, respectively. This table can be directly compared with the corresponding `Clima` polynomial (R. K. Kopparapu et al. 2013, Table 3).

Finally, we examine the dependence of these limits on planetary mass by calculating the HZ boundaries for planets of 0.1, 1, and 5 $M_\oplus$ (Figure 8). We follow the approach of R. K. Kopparapu et al. (2014), which uses the mass–radius relation $M_p/M_\oplus = 0.968\,(R_p/R_\oplus)^{3.2}$ for $M_p < 5\,M_\oplus$ and assumes that the background $N_2$ pressure scales with planetary radius as $P_s/P_\oplus = 0.937\,(R_p/R_\oplus)^{2.40}$. These relationships can be expressed using powers of mass, as $g \propto M^{1-2/3.2} = M^{0.375}$ and $p_{N_2} \propto M^{2.40/3.2} = M^{0.75}$. We renormalize both $g$ and $p_{N_2}$ so that 1 $M_\oplus$ gives exactly Earth values, with the renormalizing applied at all masses. (Without renormalizing, 1 $M_\oplus$ would give $g = 9.61$ m s$^{-2}$ and $p_{N_2} = 0.96$ bar.) The three planetary masses considered correspond to surface gravities of 4.14, 9.81, and 17.93 m s$^{-2}$ and $N_2$ inventories of 0.18, 1.0, and 3.34 bar. A more massive planet has both a higher surface gravity and a larger $N_2$ column, which reduces the water vapor

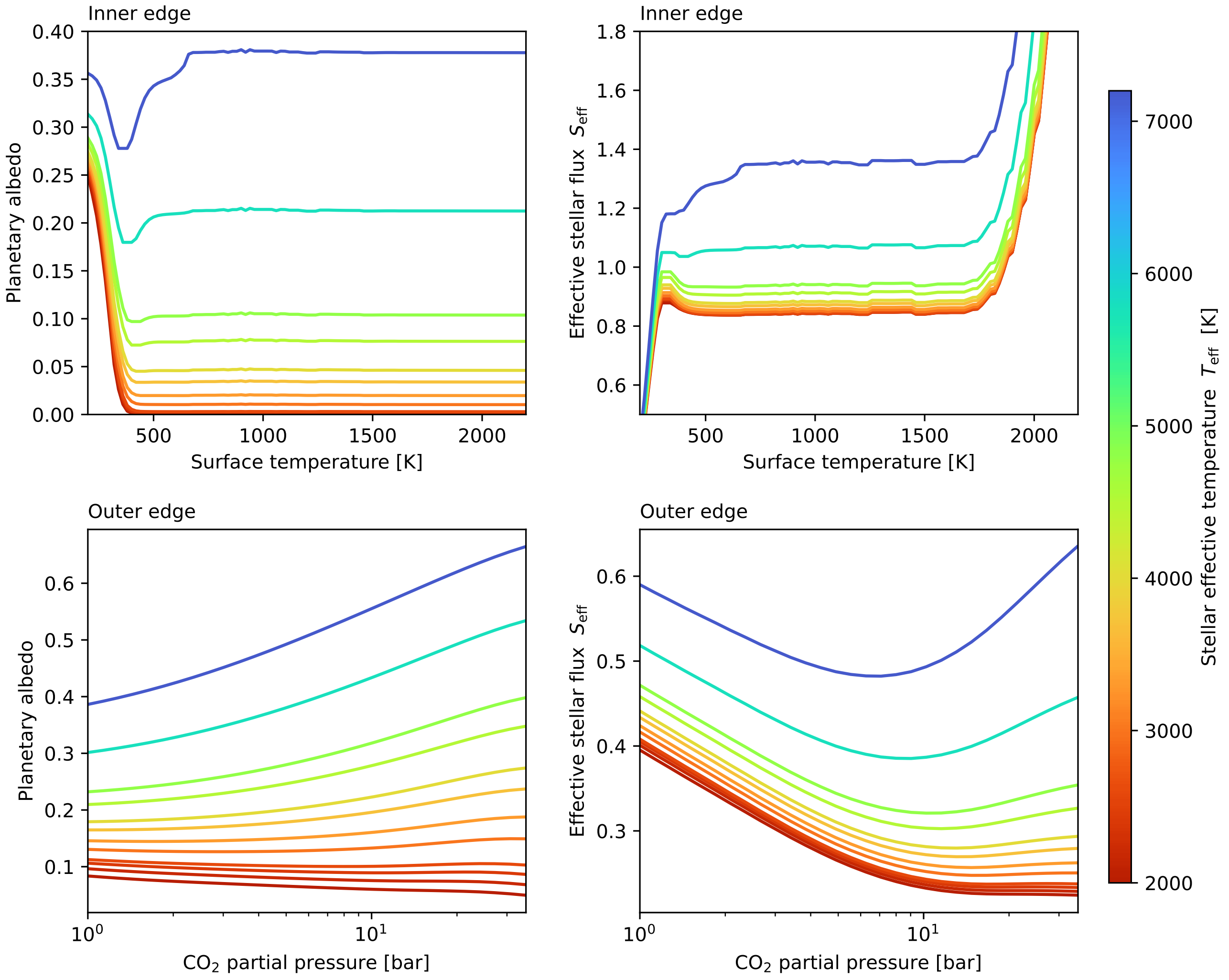

**Figure 6.** Habitable zone limits for 12 host stars ranging from M (2000 K) to F (7200 K). Inner-edge calculations are shown on the top row and outer edge calculations are shown on the bottom row.

**Table 1** Habitable zone boundary coefficients

| HZ Boundary | $S_{eff,\odot}$ | $a$ | $b$ | $c$ | $d$ |
|---|---|---|---|---|---|
| Runaway greenhouse | 1.0684 | $9.1544 \times 10^{-5}$ | $1.6309 \times 10^{-8}$ | $1.7962 \times 10^{-12}$ | $1.2497 \times 10^{-16}$ |
| Moist greenhouse | 1.0636 | $8.7628 \times 10^{-5}$ | $5.7056 \times 10^{-9}$ | $-3.2355 \times 10^{-12}$ | $-5.3193 \times 10^{-16}$ |
| Maximum greenhouse | 0.3851 | $7.1906 \times 10^{-5}$ | $3.2723 \times 10^{-9}$ | $-3.2530 \times 10^{-12}$ | $-5.4811 \times 10^{-16}$ |

**Note.** ExoColumn habitable zone boundary coefficients to calculate stellar effective flux by Equation (4) and orbital distance by Equation (5) for main-sequence stars with effective temperatures from 2000 K to 7200 K.

column depth and shifts the inner edge of the HZ closer to the host star.

Planetary mass enters the inverse calculation because the mass of a saturated column scales with the water vapor partial pressure $e_{sat}(T_s)$ as $e_{sat}/g$, so a higher surface gravity gives a smaller infrared optical depth. At $T_s = 325$ K, near the Simpson–Nakajima limit for Earth, this corresponds to about 3.3, 1.4, and 0.8 m of liquid water spread over the surface for the 0.1, 1, and 5 $M_\oplus$ planets, respectively. As discussed in Section 3, these amounts are small compared with the total water inventory, so the runaway greenhouse limit does not depend on how the ocean inventory is assumed to scale with planetary mass. We keep the 270 bar inventory at all three masses; for comparison, an Earth-like water mass fraction would correspond to about 50 bar on a 0.1 $M_\oplus$ planet and 900 bar on a 5 $M_\oplus$ planet, both of which greatly exceed the water needed to reach the limit. The outer edge remains relatively insensitive to changes in mass.

Figure 8 shows the runaway greenhouse limit at the inner edge and the maximum greenhouse at the outer edge, each drawn for 0.1, 1, and 5 $M_\oplus$. For a Sun-like host the three masses show the inner edge at $S_{eff} = 0.998$, 1.072, and 1.141, while the three outer-edge curves are identically $S_{eff} = 0.385$. The outer edge does show a modest mass dependence for cooler host stars but remains a weaker mass dependence than the inner edge. The dashed curves in Figure 8 show the comparable Clima boundaries for the dependence of the HZ on planetary mass (R. K. Kopparapu et al. 2014, Figure 3). For the Sun, the 1 $M_\oplus$ case gives $S_{eff} = 1.072$ for `ExoColumn` compared to 1.107 for `Clima`, which is a 3% difference. The 0.1 $M_\oplus$ case gives better than 1% agreement ($S_{eff} = 0.998$ versus 0.990), while the 5 $M_\oplus$ case differs by 4% ($S_{eff} = 1.141$ versus 1.188). The mass ordering is likewise consistent between the models, with higher gravity yielding a smaller atmospheric column that shifts the HZ inward. However, the solutions with `ExoColumn` at different planetary masses occupy a somewhat compressed range compared to `Clima`. This again is primarily due to the differences in the correlated-$k$ radiative transfer cores (Section 2.2).

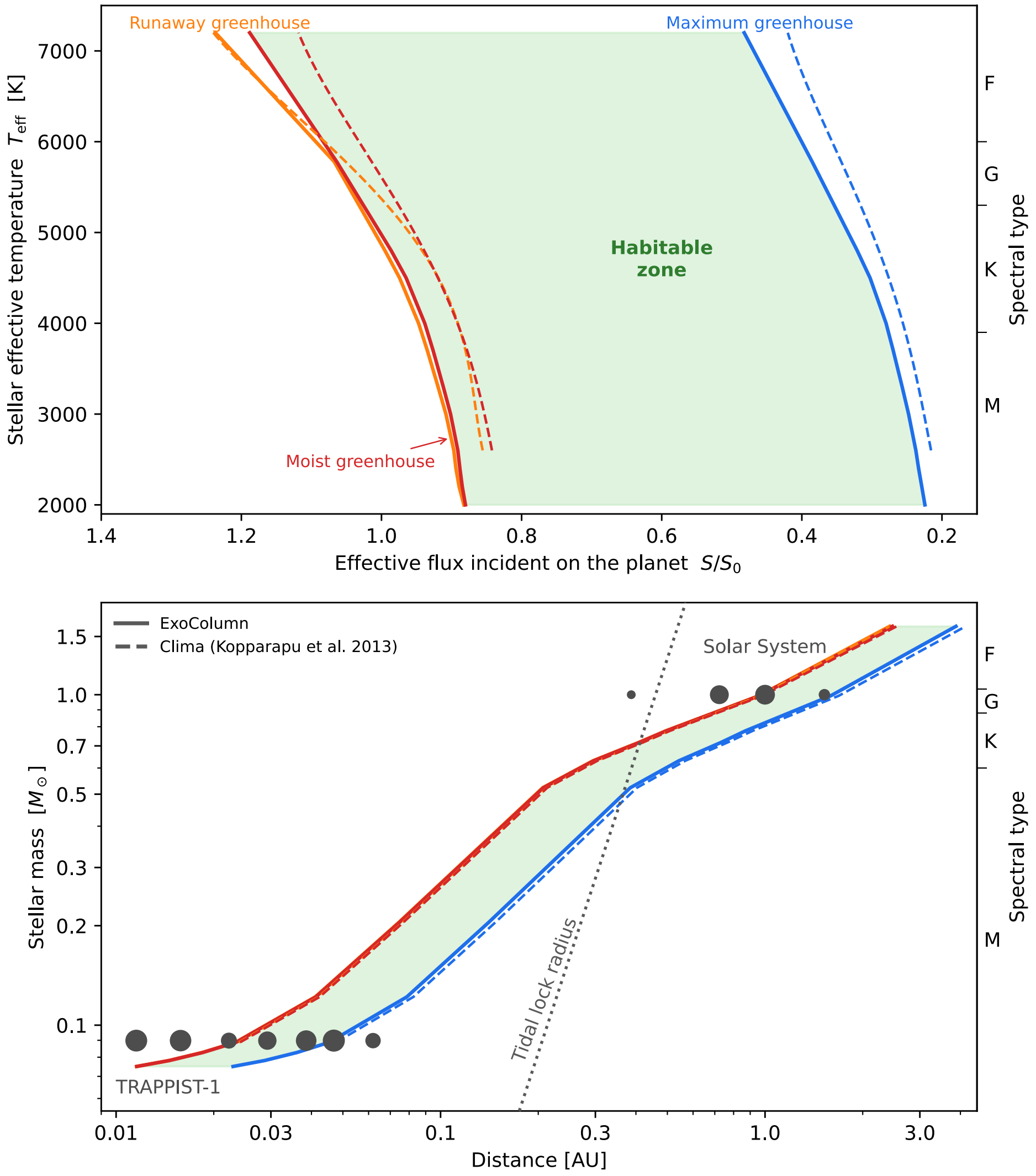

**Figure 7.** Habitable zone boundaries for main-sequence stars. The top panel shows $S_{eff}$ at the moist greenhouse (red), runaway greenhouse (orange), and maximum greenhouse (blue) limits vs. $T_{eff}$. The bottom panel shows the corresponding orbital distance vs. stellar mass, with the tidal lock radius (dotted line). The solar system and TRAPPIST-1 planets are shown for reference, and dashed curves show the R. K. Kopparapu et al. (2013) boundaries.

In summary, `ExoColumn` is capable of reproducing all HZ calculations that have been performed with `Clima`. The differences in the model results can be understood as a result of differences in radiative transfer, and even with these differences the results differ by a margin of a few percent.

## 6. Discussion and Conclusion

We have presented `ExoColumn` as a new one-dimensional RCE model with all the physical capabilities needed to perform HZ calculations comparable to those with `Clima`. The value in constructing this new model is that it leverages the existing `ExoRT` radiative transfer core, also used in the `ExoCAM` three-dimensional model, which will enable self-consistent comparisons between these two models with different dimensionality but the exact same radiative transfer. `ExoColumn` is also freely available for community use and will be included in ongoing exoplanetary model intercomparison projects (L. E. Sohl et al. 2024).

These calculations also provide an independent validation of the `Clima` HZ boundaries (J. F. Kasting et al. 1993; R. K. Kopparapu et al. 2013, 2014). `Clima` has remained an influential reference in astrobiology for over three decades and continues to be used today. During this time, three-dimensional models have also provided constraints on HZ limits, with the most significant contributions in the M-dwarf regime where synchronous rotation can significantly change the results obtained by one-dimensional models (see, e.g., A. L. Shields et al. 2016, for a review). However, it remains challenging for any single three-dimensional model to comprehensively explore the inner and outer HZ boundaries for a range of stellar types as was done with `Clima`. This study with `ExoColumn` represents the first comprehensive attempt at validating the Clima results using a model of a different lineage. The results shown in this paper are a testament to the robustness of Clima in its original (J. F. Kasting et al. 1993) and updated (R. K. Kopparapu et al. 2013, 2014) forms.

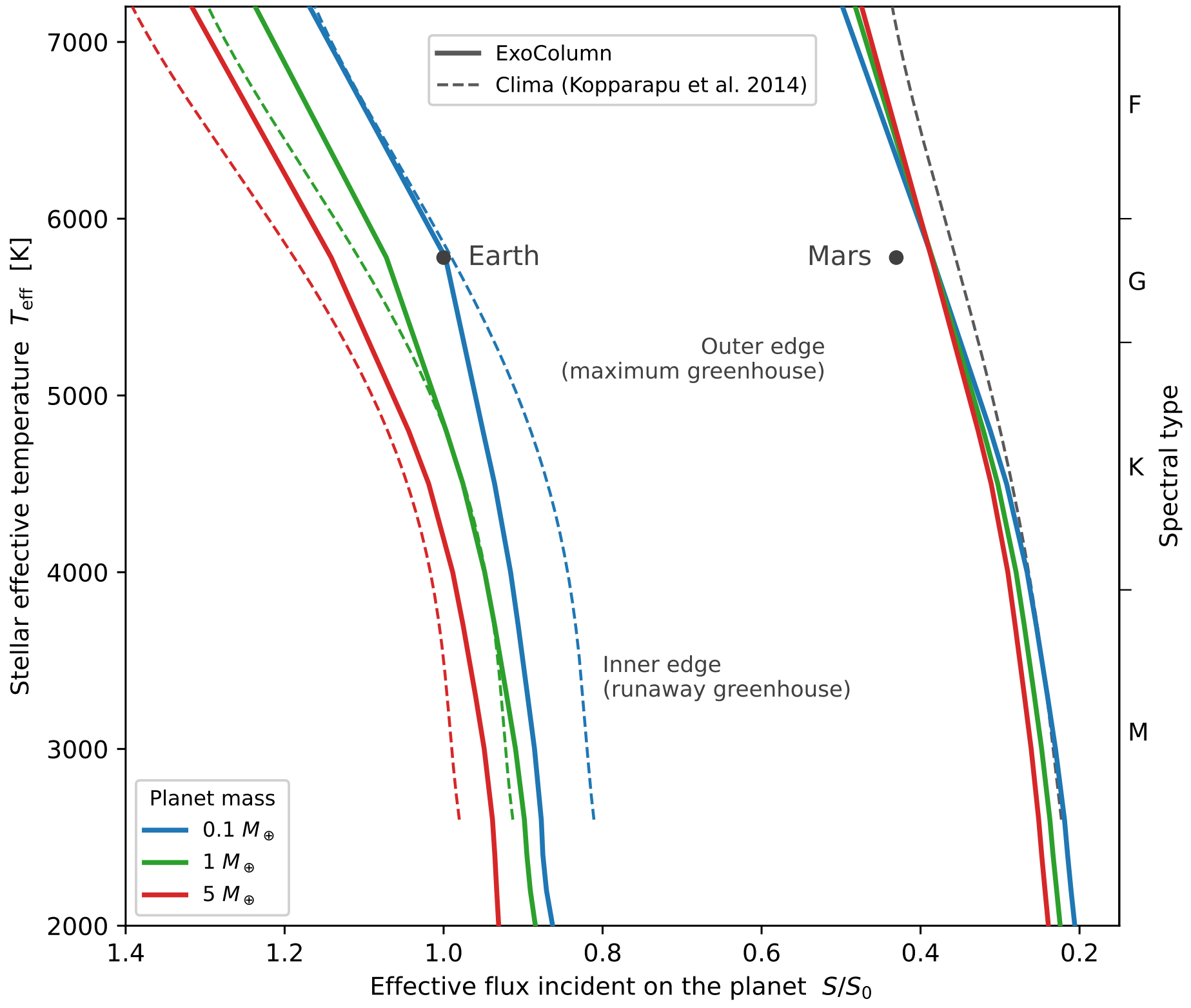

**Figure 8.** Dependence of habitable zone boundaries on planetary mass. The inner edge is the runaway greenhouse limit and the outer edge is the maximum greenhouse limit, shown for 0.1, 1, and 5 $M_\oplus$ planets. Both edges are color coded by planetary mass. Dashed curves show the R. K. Kopparapu et al. (2014, Equation (4)) boundaries.

## Acknowledgments

This paper is dedicated to Jim Kasting. This material is based on work performed as part of the Consortium on Habitability and Atmospheres of M dwarf Planets (CHAMPs) team, supported by the National Aeronautics and Space Administration (NASA) under grant No. 80NSSC23K1399 issued through the Interdisciplinary Consortia for Astrobiology Research (ICAR) program.

*Software:* ExoColumn v1.1.0 (J. Haqq-Misra 2026), ExoRT (E. T. Wolf et al. 2022), Clima (R. K. Kopparapu et al. 2013, 2014), konrad (L. Kluft et al. 2019), RADIS (E. Pannier & C. O. Laux 2019), Matplotlib (J. D. Hunter 2007), Claude Code (Opus 4.8, Anthropic 2026).

## ORCID iDs

Jacob Haqq-Misra https://orcid.org/0000-0003-4346-2611
Eric T. Wolf https://orcid.org/0000-0002-7188-1648
Ravi K. Kopparapu https://orcid.org/0000-0002-5893-2471